\pdfoutput=1
\documentclass[a4paper,11pt]{article}
\usepackage{jheppub}
\usepackage{textcase}
\usepackage{graphicx,epsfig,psfrag,amssymb,hyperref}
\usepackage{multirow}
\usepackage{color,graphicx,epsfig,psfrag,amsmath,empheq}
\usepackage[dvipsnames]{xcolor}
\usepackage{bm}
\usepackage{mathrsfs,amsfonts,soul,color}
\usepackage[caption=false]{subfig}
\usepackage{hepunits}
\usepackage{enumitem}
\usepackage{placeins}
\usepackage{mathtools}
\usepackage{siunitx}
\usepackage{textcomp}
\usepackage{tcolorbox}
\usepackage{float}
\usepackage[toc,page]{appendix}
\usepackage[ruled,vlined]{algorithm2e}
\usepackage{wrapfig}

\newcommand{\pt}{\ensuremath{p_{\text{T}}}\xspace}

\begin{document}

\title{Jets and Jet Substructure at Future Colliders}

\author[1]{Johan Bonilla,}
\author[2]{Grigorios Chachamis,}
\author[3]{Barry M. Dillon,}
\author[4]{Sergei V.~Chekanov}
\author[1]{Robin Erbacher,}
\author[5]{Loukas Gouskos,}
\author[6]{Andreas Hinzmann,}
\author[7]{Stefan H{\"o}che,}
\author[8]{B. Todd Huffman,}
\author[9]{Ashutosh. V.~Kotwal}
\author[10]{Deepak Kar,}
\author[11]{Roman Kogler,}
\author[12]{Clemens Lange,}
\author[5]{Matt LeBlanc,}
\author[13]{Roy Lemmon,}
\author[14]{Christine McLean,}
\author[15,*]{Benjamin Nachman,}
\emailAdd{bpnachman@lbl.gov}
\author[16]{Mark S. Neubauer,}
\author[3]{Tilman Plehn,}
\author[14,*]{Salvatore Rappoccio,}
\emailAdd{srrappoc@buffalo.edu}
\author[17]{Debarati Roy,}
\author[18]{Jennifer Roloff,}
\author[19]{Giordon Stark,} 
\author[7,*]{Nhan Tran,}
\emailAdd{ntran@fnal.gov}
\author[20]{Marcel Vos}
\author[21]{Chih-Hsiang Yeh}
\author[21]{Shin-Shan Yu}

\affiliation[*]{Editor}
\affiliation[1]{University of California at Davis, Davis, CA 95616, USA}
\affiliation[2]{Laborat{\' o}rio de Instrumenta\c{c}{\~ a}o e F{\' \i}sica Experimental de Part{\' \i}culas (LIP), Lisboa, Portugal}
\affiliation[3]{Universit\"at Heidelberg, Heidelberg, Germany}
\affiliation[4]{HEP Division, Argonne National Laboratory, 9700 S.~Cass Avenue, Lemont, IL 60439, USA.}
\affiliation[5]{Experimental Physics Department, Organisation Européenne pour la Recherche Nucléaire (CERN), F-01631 Prévessin Cedex, France -- CH-1211 Genève 23, Geneva, Switzerland}
\affiliation[6]{University of Hamburg, Hamburg, Germany}
\affiliation[7]{Fermi National Accelerator Laboratory, Batavia, IL 60510, USA}
\affiliation[8]{Oxford University, Oxford, United Kingdom}
\affiliation[9]{Department of Physics, Duke University, Durham, NC 27708, USA}
\affiliation[10]{University of Witwatersrand, Johannesburg, South Africa}
\affiliation[11]{Deutsches Elektronen-Synchrotron, DESY, Hamburg, Germany}
\affiliation[12]{Paul Scherrer Institute, Villigen, Switzerland}
\affiliation[13]{Daresbury Laboratory, Warrington, Cheshire, WA44AD, United Kingdom}
\affiliation[14]{University at Buffalo, State University of New York, Amherst, NY 14221, USA}
\affiliation[15]{Physics Division, Lawrence Berkeley National Laboratory, Berkeley, CA 94720, USA}
\affiliation[16]{Department of Physics, University of Illinois at Urbana-Champaign, Urbana, IL 61801, USA}
\affiliation[17]{Amity Institute of Applied Sciences, Amity University Uttar Pradesh, Noida, 201313, India}
\affiliation[18]{Brookhaven National Laboratory, Upton, NY 11973, USA}
\affiliation[19]{Santa Cruz Institute for Particle Physics, UC Santa Cruz, CA 95060, USA}
\affiliation[20]{IFIC (UV/CSIC) Valencia, 46980 Paterna, Spain}
\affiliation[21]{Department of Physics and Center for High Energy and High Field Physics, National Central University, Chung-Li, Taoyuan City 32001, Taiwan}

\abstract{
Even though jet substructure was not an original design consideration for the Large Hadron Collider (LHC) experiments, it has emerged as an essential tool for the current physics program.  We examine the role of jet substructure on the motivation for and design of future energy frontier colliders.  In particular, we discuss the need for a vibrant theory and experimental research and development program to extend jet substructure physics into the new regimes probed by future colliders.  Jet substructure has organically evolved with a close connection between theorists and experimentalists and has catalyzed exciting innovations in both communities.  We expect such developments will play an important role in the future energy frontier physics program. 

}

\maketitle

\section{Introduction}

Jets produced from high energy quarks and gluons through quantum chromodynamics (QCD) have a complex composition.  This jet substructure (JSS) has emerged as a powerful framework for studying the Standard  Model (SM) at particle colliders, and provides a key set of tools for probing nature at the highest energy scales accessible by terrestrial experiments~\cite{Abdesselam:2010pt,Altheimer:2012mn,Altheimer:2013yza,Adams:2015hiv,Larkoski:2017jix,Kogler:2018hem,Marzani:2019hun,Kogler:2021kkw}.  
While not an experimental or theoretical consideration of the Large Hadron Collider (LHC) experiments' original designs, JSS is now being widely used to extend the sensitivity of searches for new particles, to enhance the precision of measurements of highly-Lorentz-boosted SM particles, as well as to probe the fundamental and emergent properties of the strong force in new ways.  Along the way, the JSS community has been a catalyst for new detector concepts, new analysis tools (e.g., using deep learning), new theory techniques, and more.  Jet substructure has transformed the physics program of the LHC and it can play a central role in the physics case for and the design considerations of future colliders. 

As the particle physics community decides what the direction of the field should be in the middle part of the 21$^\text{st}$ century, it is useful to assess the state of JSS techniques that have developed over the last decades and to highlight the utility in various future collider scenarios. Efforts to investigate these scenarios are currently under way by the broader community, with pros and cons for many different strategies, for instance in the European Committee for Future Accelerators~\cite{Detector:2784893}, and as part of the Snowmass 2021 process in the US (for which this paper is a contribution)~\cite{snowmass}. While it is not yet clear what the future energy frontier machine(s) will be, it is clear that jets and JSS will play an important role in the physics program of the future. 

In this forward-looking perspective paper\footnote{This paper is not a review and is not comprehensive.  See the reviews cited earlier for an in depth view of the state of JSS.}, we will investigate the opportunities and challenges associated with the various types of future colliders in the context of JSS. We will discuss both lepton and hadron colliders, including Higgs factories and ultra high energy machines.  This paper is organized as follows.  In Sec.~\ref{sec:sigs}, we give a brief introduction to various signatures of interest for JSS physics.  We then outline the multiple avenues of research that the will be important in the context of Snowmass 2021. We believe that numerous topics of relevance for the Snowmass process should be discussed and evaluated with explicit considerations of the impacts for and benefits from JSS theory, phenomenology, and experimental tools (both hardware and software). These topics will be covered in a section on Theoretical Innovation (Sec.~\ref{sec:theoretical_innovation}), Experimental Innovation (Sec.~\ref{sec:expinnov}), and Enhancing Sensitivity (Sec.~\ref{sec:enhance}).  We forgo a conclusion section in favor of the executive summary preceding this introduction.


\section{Signatures of Interest}
\label{sec:sigs}

There are a large number of signatures that can benefit from JSS at future colliders.  In general, JSS techniques are applied to tag Lorentz-boosted massive particles ($H/W/Z$ bosons, top quarks, and BSM particles) and to explore the structure of the strong force in final state radiation on small angular scales.  This section briefly introduces various categories in the context of both SM measurements and BSM searches.


\subsection{Light Quark and Gluon Jets}

\noindent \textbf{Measurements}: High energy quark and gluon jets provide important probes of a variety of quantum chromodynamic (QCD) phenomena.  These jets can be used to study perturbative aspects of QCD as well as features of QCD that cannot currently be described with perturbation theory.  For the latter case, there are cases where scaling relations can be predicated and tested across a wide range of energies.  These final states can be used to measure the strong coupling constant, to extract various universal objects within factorized QCD, to tune Parton Shower Monte Carlo generators, as well as other tasks.  Quark and gluon jets were also studied at previous colliders, but higher energy machines allow for a suppression of non-perturbative effects as well as a larger lever arm for testing scaling behaviors. \\

\noindent \textbf{Searches}: Quark and gluon jets are statistically distinguishable due to their different fragmentation processes.  Quark versus gluon jet tagging has been a standard benchmark for the development of new classical and machine learning-based jet taggers.  Many SM and BSM final states of interest are dominated either by quark or gluon jets, in contrast to the dominant background processes.  Quark versus gluon jet tagging can help enhance such signals, although these jets are not as seperable as other objects described below.

\noindent See also references~\cite{Gras:2017jty,Proceedings:2018jsb,Amoroso:2020lgh} for further details.

\subsection{Bottom Quarks}

Bottom quarks are prevalent in BSM decays as well as in the decays of $H/Z$ bosons, and top quarks.  Bottom quark jets are highly separable from other jets due to the long lifetime of the bottom quark and the heavy mass of bottom-flavored hadrons.  In addition to lifetime information, jet substructure can be used to further separate these jets from other jets~\cite{CMS:2017wtu,ATLAS:2019bwq}.  

A similar story is true to a lesser extent for charm quark jets~\cite{CMS:2021scf,ATLAS:2021cxe} and to an even lesser extent for strange quark jets~\cite{Erdmann:2020ovh,Nakai:2020kuu,Erdmann:2019blf}. 


\subsection{$H$ boson}

A main goal of the HL-LHC, as well as future Higgs factories, is to study the $H$ 
boson~\cite{ATLAS:2012yve,CMS:2012qbp} in as much detail as possible. This includes
detailed measurements of the branching fractions (BF). In the $H\rightarrow b\overline{b}$ and $H\rightarrow c\overline{c}$ final states, current analyses at 
the LHC~\cite{ATLAS:2018kot,CMS:2018nsn,ATLAS:2020fcp,ATLAS:2020jwz,ATLAS:2016mzy,CMS:2015ebl,ATLAS:2020bhl} utilize kinematic criteria that identify hadronically 
decaying $H$ bosons that have Lorentz factors larger than 1. These are moderately to 
fully boosted topologies. 
identify these final states. 
In addition, BSM physics that decay to $H$ bosons (or other Higgs-like scalars) can 
also utilize these reconstruction techniques as is done in the current LHC 
experiments (a review can be found in Ref.~\cite{GOUZEVITCH2020100039}). In particular, specifically for bottom and charm quark final 
states, flavor and lifetime information can be used in addition to the jet substructure to improve categorization. Many all future collider scenarios result in copious Higgs bosons produced with large Lorentz boosts, so the techniques developed at the LHC will be broadly applicable for these cases as well. 

\subsection{$W/Z$ bosons}

The cross sections and branching ratios of $W$ and $Z$ bosons are extremely well known via leptonic channels and previous LEP measurements~\cite{ParticleDataGroup:2020ssz}. However, $W$ and $Z$ bosons often participate in BSM scenarios, so can be present in many final states of these models (see a review in Ref.~\cite{RAPPOCCIO2019100027}). For example, in SM extensions with an additional real~\cite{vonBuddenbrock:2016rmr} or complex~\cite{PhysRevD.97.015022} scalar field $S$, resonant $hS$ and $SS$ production~\cite{Robens:2019kga,Dawson:2017jja,Muhlleitner:2017dkd} can lead to an enhanced rate of highly-boosted $W/Z$ bosons.

The identification of $W$ and $Z$ bosons is similar to the $H$ boson, however the masses are slightly lower and they often do not decay to bottom or charm quarks, so there are fewer handles to use to identify them.


However, the phenomena of $W$ and $Z$ bosons radiating off of very high energy jets (``Weak-strahlung'') is 
a new area that is not very likely at the LHC (although there are measurements of $W$ 
bosons nearby jets~\cite{ATLAS:2016jbu}). Even standard QCD jets that originate from 
quarks or gluons can have additional information from jet substructure. It provides an 
opportunity to study the phenomenon. 


In addition, $W/Z$ bosons in vector-boson fusion initial states can also be more highly boosted than in $s-$ or $t-$ channel creation. Boosted techniques can also be used to appropriately identify these collisions at future colliders. 

Finally, study of the vector boson scattering (VBS) process informs the degree to which the Higgs mechanism is the source of EWSB and thus provides an important test of the SM. Additionally, new physics that alters the quartic gauge couplings (QGC)~\cite{Eboli:2003nq,Eboli:2006wa}, or involves new resonances~\cite{Chang:2013aya,Espriu:2012ih}, predict enhancements for VBS at high $p_{\rm{T}}$ of the vector bosons and invariant mass of the diboson system. At large $m(VV)$ of most interest, two $W/Z$ bosons are produced with large momentum and a hadronically-decaying boson can be reconstructed using boosted-boson tagging techniques that exploit jet substructure~\cite{Cacciari:2008gp,Krohn:2009th,Ellis:2009su,Butterworth:2008iy,Aad:2013gja}.

\subsection{Top quark}

The top quark is a special quark with a Yukawa coupling close to unity. This makes it a likely participant in many BSM models to explain the hierarchy problem. The top quark nearly always decays to a $W$ boson and a bottom quark~\cite{ParticleDataGroup:2020ssz}. At the LHC, even SM production of top quark pairs can result often in boosted final states~\cite{CMS:2016poo,CMS:2017pcy,CMS:2019fak,CMS:2020tvq,CMS:2021vhb,ATLAS:2022xfj}. In addition, many BSM scenarios have boosted top quarks participating in the event (a review can be found in Ref.~\cite{RAPPOCCIO2019100027}). 

The jet substructure of top quarks is, in some sense, an ideal case, since there are two heavy SM particle masses to utilize (the top quark and $W$ boson), as well as lifetime and flavor information in the final state particles. This provides a strong handle to identify top quarks. 

Especially at higher-energy future colliders, the analysis of collisions containing top quarks will be ever more reliant on jet substructure and boosted topologies. Similarly to the $W$ and $Z$ bosons, there may also be top quark production within a jet that originates from light quarks or gluons via gluon splitting to $t\overline{t}$, similarly to the case at the Tevatron and LHC for bottom quarks. These types of events will need to be handled separately from events without these gluon splittings. Jet substructure and boosted techniques will play an increasingly important role here also.  \\

\noindent\textbf{Multi-class Tagging:} While most tagging studies focus on binary classification (one signal versus one combined background), it is also possible to simultaneously tag multiple signals at the same time (see e.g. Ref.~\cite{JME-18-002}).  Multiclass classification methods output a score for each signal and background type that often corresponds to the probability that the jet belongs to the class given the inputs (with prior probabilities as in the dataset).  While such approaches may not necessarily improve classification accuracy (with sufficient training examples), they can provide flexibility for downstream analyses.


\subsection{Background Processes}

For all of the signatures described above, there are a variety of physics backgrounds that obfuscate the target signatures.  At hadron colliders, this is the result of multiple, nearly simultaneous collisions (pileup) as well as underlying event, and multiparton interactions.  A variety of \textit{jet grooming} techniques have been developed to mitigate these effects (see e.g., Refs.~\cite{Abdesselam:2010pt,Altheimer:2012mn,Altheimer:2013yza,Adams:2015hiv}).  While similar backgrounds in $e^+e^-$ are often much smaller, beam-induced backgrounds in muon colliders~\cite{Collamati:2021sbv} could potentially benefit from similar techniques developed for hadron colliders.

\section{Theoretical Innovation}
\label{sec:theoretical_innovation}

In the last several decades, major advances in theoretical techniques have drastically improved our understanding of the nature of QCD radiation (a review is found in Ref.~\cite{Larkoski:2017jix}). A combination of fixed-order, resummation, non-perturbative, and machine-learning techniques have opened new avenues of study, guided by extensive measurements of these processes at the LHC and elsewhere. Some studies of these topics with respect to collider scenarios is highlighted in Ref.~\cite{Amoroso:2020lgh}. In this Section, we focus on the develompents of Monte-Carlo (MC) event generators, particularly as applied to new collider scenarios and to improve JSS modeling.



\label{sec:mctuning}
Monte-Carlo (MC) event generators provide the link between the theoretical calculations and experimental measurements through a fully differential simulation of final states~\cite{Buckley:2011ms}. These are a combination of fixed-order, resummed, and non-perturbative effects~\cite{Berger:2008sj,Bevilacqua:2011xh,Cascioli:2011va,Hirschi:2011pa,Cullen:2014yla,Actis:2016mpe}. At present, the majority of the uncertainty that lies in JSS is in the so-called ``physics model''~\cite{ATLAS:2017zda,CMS:2018vzn}, which includes the parton shower and hadronization, the former of which performs the QCD evolution, and the latter of which is performed with either the Lund string model~\cite{Andersson:1983jt,Andersson:1983ia} or the cluster model~\cite{Gottschalk:1982yt,Gottschalk:1983fm,Webber:1983if}. 

The MC event generators most commonly used to compare to experimental measurements at the LHC are Herwig~\cite{Bellm:2015jjp}, Pythia~\cite{Sjostrand:2014zea} and Sherpa~\cite{Sherpa:2019gpd}. They contain various parton-shower models for the simulation of jet evolution, and cover a broad spectrum of matching and merging techniques. Several recent studies compared the physics performance of these generators for a large number of processes of relevance to the LHC~\cite{Kanaki:2000ey,Krauss:2001iv,Mangano:2002ea,Gleisberg:2008fv,Alwall:2014hca,Andersen:2016qtm,Gras:2017jty,Bellm:2019yyh,Buckley:2021gfw} and observed good agreement in their predictions for identical input parameters. For any given generator, the prediction may however strongly depend on those parameters, i.e., on the generator tune. Improvements in these tools will give a better event-by-event simulation of collisions with JSS, and will allow better modeling of background processes as well as better inputs to advanced ML-based techniques. 



One typical parametric uncertainty is the value of the strong coupling. Another common systematic uncertainty is the recoil scheme in the parton shower, which impacts a Monte-Carlo prediction in a different way than an analytical resummation, due to momentum and probability conserving effects in the event generator. These effects must however not influence the Monte-Carlo result in those regions where momentum conservation becomes irrelevant, and where analytic results can be obtained for certain observables. Much effort has been devoted recently to understanding these constraints in the context of parton-shower algorithms~\cite{Hoche:2017kst,Dasgupta:2018nvj,Nagy:2020rmk}, and in providing parton showers that satisfy the theoretical boundary conditions~\cite{Nagy:2007ty,Bewick:2019rbu,Dasgupta:2020fwr}. In addition, some observables require the understanding of sub-leading color and spin effects, which are typically absent in parton-showers used for LHC physics. There has been renewed interest in implementing algorithms to include these spin correlations~\cite{Knowles:1988hu,Nagy:2008eq,Richardson:2018pvo,Karlberg:2021kwr,Hamilton:2021dyz}, and in including sub-leading color corrections for non-global observables~\cite{Nagy:2019pjp,Hamilton:2020rcu}.
Some efforts have also been made to devise a generic approach for implementing higher-order corrections to the parton-shower splitting kernels in a fully differential form~\cite{Hartgring:2013jma,Li:2016yez,Hoche:2017iem,Dulat:2018vuy,Gellersen:2021eci}. All these improvements will help to link analytic predictions for resummed jet observables to event generator predictions.

Systematic uncertainties also arise in the combination of fixed-order computations with parton showers. Matching algorithms for next-to-leading order (NLO) QCD calculations~\cite{Frixione:2002ik,Nason:2004rx} mainly differ in their treatment of real-radiative corrections. When observables become sensitive to either radiation (e.g.\ jet-$p_T$) or inhibited radiation (e.g.\ jet veto), this difference can create the dominant uncertainty. Similarly, merging algorithms, both at leading order~\cite{Catani:2001cc,Mangano:2001xp,Lonnblad:2001iq} and at next-to-leading order~\cite{Hoeche:2012yf,Frederix:2012ps,Lonnblad:2012ix} have associated uncertainties, which are mostly related to the matching algorithm for the underlying NLO calculations, and to the treatment of unitarity~\cite{Lonnblad:2012ix,Bellm:2017ktr}. Uncertainties in current NNLO matching algorithms arise from the precise technique being used to devise the resummed result at small transverse momentum in the case of resummation based approaches~\cite{Hamilton:2012rf,Alioli:2013hqa,Monni:2019whf}, and again from the treatment of unitarity in all approaches~\cite{Lonnblad:2012ix,Hoche:2014uhw,Bellm:2017ktr,Prestel:2021vww,Bertone:2022hig}.

Finally, systematic undertainties may arise from the implementation of semi-hard physics effects, such as multiple scattering~\cite{Sjostrand:1987su,Butterworth:1996zw} and hadronization~\cite{Bengtsson:1987kr,Andersson:1997xwk,Webber:1983if,Winter:2003tt}. It is to be kept in mind that often the study of hadronization uncertainties is performed by replacing not only the hadronization model itself, but also the parton shower. This procedure is ill-advised, as the true hadronization uncertainty is almost always overestimated (however, see Ref.~\cite{Ghosh:2021hrh}). Studies using different hadronization models with identical perturbative input have demonstrated that in many cases the hadronization uncertainties are subdominant~\cite{Hoeche:2011fd,SM:2012sed,Bellm:2019yyh}.

\section{Experimental Innovation}
\label{sec:expinnov}

Future colliders often have substantively different characteristics compared to the LHC. Higher-energy $pp$ colliders will have more radiation and pileup, with the SM particles being produced with enormous Lorentz boosts and often in the forward region of the detector. Muon colliders will have beam-induced backgrounds. Electron-positron colliders have simpler environments due to lack of pileup and a precise measure of the $z$ position of interactions. These all come with challenges and opportunities that can be exploited. This can come in the form of detector optimization for JSS, improved reconstruction algorithms, and in improved calibration and systematic uncertainties. These are covered in the following section.  

\subsection{Detector Optimization}
\label{sec:detopt}

There are several detector technologies that will improve JSS and related techniques. These include finer calorimeter granularity~\cite{Yeh:2019xbj,Coleman:2017fiq}, more hermetic coverage of tracking detectors, and precise measurements of timing information. The experience of the LHC has shown that such information can be used to more accurately reconstruct the interaction of hadrons with various detector elements, much of which is used in the `particle flow' (PF) concept salready deployed by the LHC experiments. At future muon colliders, `beam background' detectors could also in principle be deployed to reduce the impact on JSS.

\subsubsection{Electron-positron Colliders}

The main detector concepts developed for electron-positron collider experiments are based on PF. With transparent, hermetic trackers and highly granular calorimeters, the ILD~\cite{ILD:2019kmq} and SiD~\cite{Breidenbach:2021sdo} experiments at the ILC, as well as the CLIC detector~\cite{CLICdp:2018vnx} and the CLD design~\cite{Bacchetta:2019fmz} for the FCC-$ee$, are designed to efficiently associate tracks and calorimeter energy deposits. A global detector R\&D program has proven the feasibility of highly granular calorimeters~\cite{CALICE:2011brp} and large-scale systems are under construction for the ALICE~\cite{ALICECollaboration:2719928}, ATLAS~\cite{CERN-LHCC-2015-020}, and CMS~\cite{Contardo:2020886} upgrades. The optimization of the overall design was primarily driven by the jet energy resolution, but as a collateral benefit, these concepts offer excellent substructure performance. Jet substructure studies based on full simulation have been performed in Ref.~\cite{Strom:2020wrm}. 




In addition to the intrinsic particle
identification capabilities, the fine transverse granularity allows close showers to be separated and provides good matching to tracks in the inner preshower signals, and also to muon tracks, making this calorimeter a good candidate for efficient particle-flow reconstruction. The need for
disentangling signals produced by overlapping electromagnetic and hadron showers
is likely to require longitudinal segmentation as well. Several ways to implement this
segmentation were envisioned and are being studied, e.g. the classical division of the
calorimeter in several compartments, an arrangement with fibres starting at different depths, the extended use of the timing information, etc. The specific advantages
and drawbacks of each approach need to be studied through both simulations and
beam tests. High-granularity calorimetry associated with a silicon tracker will be a promising option to reach jet energy resolutions around 5-20\% with PF reconstruction.

\subsubsection{Muon Collider}

%

Proposed muon colliders offer a physics reach for discoveries similar to that of proposed high-energy hadron colliders, while maintaining appealing experimental aspects of lepton collider environments such as a lack of pileup and underlying event, as well as precise determination of the $z$ position of the interaction. A critical difference between muon and electron accelerators is the presence of large beam-induced background (BIB) processes for muon machines, which arise due to muons in the beam decaying via $\mu \rightarrow e \nu \bar{\nu}$ before colliding. The resultant electrons interact with experimental elements along the beamline, creating electromagnetic showers of soft photons and neutral particles that can interact with detectors.

Detectors at future muon colliders will need to incorporate specifically-designed shielding and subsystems to mitigate BIB processes. The exact characteristics of the BIB depend strongly on the machine centre-of-mass energy and accelerator lattice, and must be studied in-detail for different scenarios. For studies during the Snowmass 2021 community planning exercise, the performance of a modified version of the CLIC detector has been benchmarked at a $\sqrt{s}=$1.5~TeV~(3~TeV)~collider. This detector includes a modified vertex detector barrel that does not overlap with regions of large BIB activity, and shielding nozzles made of Tungsten and borated polyethlene to absorb contributions from beam-induced particles. Sets of adjacent sensors in the inner detector can also be used to mitigate contributions from BIB processes, by exploiting angular correlations as done in the CMS track trigger. The experimental conditions at a muon collider will also necessitate an increased material budget for the inner tracking systems, up to 10 times larger per-layer than that foreseen for ILC detectors due to additional cooling, power and support structures.


Early studies of this detector indicate that BIB contributions will be approximately evenly distributed in the calorimeter ($\eta-\phi$), suggesting that the advanced pileup mitigation techniques studied at the LHC could also provide a versatile handle with which to remove BIB contamination during reconstruction (sec.~\ref{sec:recoalgs}). While the jet reconstruction efficiency for early jet reconstruction approaches at future muon colliders is above 90\% for high-$p_{\text{T}}$ jets, the decreased efficiency at lower jet $p_{\text{T}}$ could also imply decreased performance when reconstruction jet substructure observables which rely on subjet identification (\emph{e.g.} $N$-subjettiness~\cite{Thaler:2010tr,Thaler:2011gf}) or soft radiation patterns (\emph{e.g.} $D_{2}$~\cite{Larkoski:2014gra}).

\subsubsection{High-energy Hadron Collider}
\label{sec:detector_100tev_pp}


There are currently two main hadron-hadron collider proposals, the FCC-hh at CERN and the SPPC in China, both targeting $pp$ collisions at a center of mass energy of about 100 TeV. Driven by the physics requirements, the 100 TeV machine will deliver an integrated luminosity of around 25 ab$^{-1}$ per experiment, reaching an instantaneous luminosity of $3\times 10^{35}$ cm$^{-2}$ $\mathrm{s}^{-1}$, almost an order of magnitude larger than expected from the HL-LHC. These are extremely ambitious projects requiring breakthroughs in accelerator technology, detector design, and physics object reconstruction, and a coherent effort in all aspects is required.
 
To meet the physics requirements, the detectors for a 100 TeV machine should be able to reconstruct multi-TeV physics objects, while in parallel provide the necessary precision to measure the SM processes which typically results in high-energy final states at very high rapidity. The detector coverage should be extended  with respect to the LHC detectors, since due to the almost a factor of 5 increase in the center of mass energy, many processes are expected to be extremely forward. For instance, SM $ZZ$ production would produce two $Z$ bosons with multi-TeV energies, with transverse momenta less than 100 GeV. These would have relativistic boosts of $\gamma = 20$, with opening angles between the $Z$ boson decay products of about $0.1$ radian. Detector capabilities to reconstruct these objects are fairly challenging (for instance, the average $Z$ boson from $ZZ$ production would shower mostly within one of the current LHC calorimeter cells). Concrete detector proposals are not yet in place, however different studies have been carried out to motivate the main aspects of the design. 

An additional challenge is that the detector design should take in to account the harsh conditions expected at a 100 TeV machine. The foreseen upgrades of the LHC experiments for HL-LHC give a useful insight of the the challenges and the technology requirements expected in a future machine.

In addition to the extremely high energies that occur at very high rapidities necessitating finely granular detector elements, one of the big challenges at 100 TeV colliders is the large pileup. At the LHC, the average pileup is around 25, and it is expected to reach values of around 150-200 during the HL-LHC operation. This will result in significant degradation in the physics object reconstruction performance and hence on the physics outcome without dedicated detector systems and reconstruction algorithms. To this end, new developments are required in both the detector and reconstruction fronts. On the detector front, ATLAS and CMS experiments are developing fast timing detectors to improve the track-to-vertex association~\cite{CERN-LHCC-2020-007,Butler:2019rpu}. These technologies achieve a timing resolution $\mathcal{O}$(30) ps and studies using simulated samples show that are able to restore the physics object reconstruction performance obtained with much smaller pileup. At a 100 TeV machine, a factor of 5 larger pileup is expected posing even stringent criteria on the detector design. Likely, the developments on the precision timing detectors towards the HL-LHC will provide a solid ground to build upon. To cope with the pileup expected at 100 TeV, the timing resolution of the detectors should be improved by around a factor of 5-6, reaching a timing resolution better than 10 ps. 

The calorimetry systems must provide excellent energy resolution over a wide range of energies in the central and forward regions, 
and increased hermetic coverage with respect to the LHC ones (reaching $|\eta|<6$). Studies have shown~\cite{calohep} that another parameter of particular importance for JSS measurements in the ultra-relativistic regime, is the granularity of the detector. These studies showed that calorimeters must have 10 times finer granularity than the ones used at the LHC to achieve similar levels of performance in the main JSS observables in the this high-$p_T$ regime. The extreme levels of radiation present in a 100 TeV collider pose another challenge for the calorimeter design. 

Technologies developed and successfully used at the LHC can serve as a promising starting point. One option for the electromagnetic and hadronic calorimeters, ECAL and HCAL, respectively can be based on the concepts used for the ATLAS calorimetry system. Their ECAL system uses Liquid Argon to generate the signal from the traversing particles. This technology provides both powerful performance together with the necessary radiation tolerance. In the case of the barrel region of HCAL, a more cost-efficient solution can be explored. For instance, the ATLAS HCAL uses organic scintillating tiles as active material. For the absorber, a combination of lead and steel provides promising results. However, due to the larger levels of radiation in the endcap and forward regions, this technology is not viable. Technologies based on liquid argon can be employed also in this case. Another option in this region could be a silicon-based or hybrid silicon/photomultiplier calorimeter similar to that being deployed by CMS in the HL-LHC upgrades, the High-Granularity Calorimeter (HGCAL)~\cite{CERN-LHCC-2017-023}. This also provides a large amount of resolution for substructure determination. This detector design also provides timing information (with $\mathcal{O}$(30)ps) allowing even for a 4D particle shower reconstruction. This approach can be powerful in  suppressing the effect from pileup in the calorimeter system, and also aid the reconstruction of exotic signatures. The energy resolution in electromagnetic showers is characterized by a stochastic term $\sim$16\%/$\sqrt{E}$. Another idea for the ECAL system  is based on monolithic active pixel sensors (MAPS). Studies in simulation using $50\times50$ $\mu$m pixels and a sensitive layer thickness of 18$\mu$m yield a stochastic term of $\sim$13\%$\sqrt{E}$~\cite{FCC:2018vvp}.

The FCC-hh collaboration developed a baseline detector based on these principles~\cite{FCC:2018vvp}. The detector concept relies heavily on the ATLAS technology for both ECAL and HCAL, however changes in the design of the detector and its granularity have been considered. For instance, to complement the tracking system in JSS, the $\eta-\phi$ granularity of ECAL (HCAL) is around $\Delta\eta \times \Delta\phi = 0.01 \times 0.009$ $(\Delta\eta \times \Delta\phi = 0.025 \times 0.025)$ in the barrel region, around four times finer compared to the LHC detector. 
This transverse granularity for the 100~TeV collision environment was determined using fast Monte Carlo simulations for boosted jets at tens-of-TeV scale \cite{2015gran1}.
Detailed studies~\cite{FCC:2018vvp,detector100TeV2017} using Full Simulation demonstrated that this technology could attain a stochastic and constant term of 8 (48)$\%/\sqrt{E}$ and 0.2 (2)\%, respectively for electromagnetic (hadronic) showers, with small dependence on $|\eta|$ and neglecting pileup interactions. This can attain jet energy resolutions of $<5$\% for jets with $p_T>1$~TeV. 
Jet substructure variables for hadronic jets from highly Lorentz-boosted weak bosons from resonances between 5-40 TeV were studied in Ref.~\cite{Yeh:2019xbj}, using several spatial sizes of calorimeter cells. The current scale of LHC cell sizes around $\Delta\eta \times \Delta\phi = 0.1\times 0.1$ were insufficient to ascertain the jet substructure. 
The study confirmed the HCAL design of the baseline
FCC-hh with $(\Delta\eta \times \Delta\phi = 0.025 \times 0.025)$.
It is interesting to note that, for very boosted jets with transverse momenta close to
20~TeV, further decrease of cell size to $(\Delta\eta \times \Delta\phi = 0.0043 \times 0.0043)$  did not
show a further improvement in performance.



\subsection{Reconstruction Algorithms}
\label{sec:recoalgs}
Reconstruction algorithms for jets and jet substructure have been widely developed in the last decade. Different collider scenarios can utilize different aspects of these advancements to address their unique challenges and opportunities as compared to the LHC. However, overall there are well-established techniques to achieve the desired performance level in all scenarios, as will be described in this section. 


\subsubsection{Jet reconstruction}

Precise and well-understood jet finding, clustering, and calibration is a key initial step to deploying powerful JSS techniques.  The conceptual task is similar at the different future colliders under consideration. However, the varying energy range, whether or not the center-of-mass is known or not, and the level of beam backgrounds play a role in the optimal approach.  
Furthermore, good jet performance is reliant on well-understood and calibrated inputs for each of the subdetector elements and at the single particle level. 

In the case of the ILD and FCC-ee detectors, software compensation has been 
shown to reduce the jet energy resolution significantly~\cite{CALICE:2012eac}. 
The lack of pileup results in smaller stochastic terms, and an optimal 
assignment of tracks and clusters in the particle flow algorithm 
can lead to superior energy resolutions. 
However, differences are present in the simulation of shower shapes, 
in particular the energy and radius of the interaction region, which 
need further studies and improved simulations~\cite{Dannheim:2019rcr}.
Detailed measurements of the spatial and temporal development of showers 
in test-beam setups with fully integrated detector prototypes will 
help to improve the systematic uncertainties in the detector simulation, 
which is a crucial ingredient for precision measurements at future colliders.

For higher energies and more granular detector technologies, some initial studies have been performed.  The energy calibration of calorimeter cells, composite clusters, single particles and jets is a challenging task at a 100~TeV $pp$ collider. First studies exist on the energy calibration of the single-particle response of a FCC-hh detector, with electronic noise added to single cells and a simulation of in-time pileup~\cite{Aleksa:2019pvl}. In this study, energy deposits inside the 
calorimeter are summed into clusters using the sliding windows algorithm. 
For an optimal single-particle response, dead material corrections and a layer 
correction, accounting for the different sampling fractions depending on the 
depth of the shower, are necessary. The achieved jet resolutions are within 
the design goals with stochastic terms below 50\%, but rely on extrapolations 
from detector simulations. Hadronic and electromagnetic shower 
components up to several TeV need to be simulated, where extrapolations to these 
high energies come with large uncertainties. Differences in the hadronic shower 
simulation models in Geant4~\cite{GEANT4:2002zbu} have been reported for pions in the energy range between 2 and 10 GeV~\cite{CALICE:2019vza}. Detailed studies at higher energies will be needed to achieve the best possible precision at future colliders. 

\paragraph{Electron-positron colliders}

Compared to the previous generation of high-energy electron-positron colliders, the complexity of final states increases considerably. However, this complexity is comparable to that already observed in the LHC experiments. For instance, the hadronic Higgs-strahlung analysis at a Higgs factory requires excellent jet clustering performance in four-jet final states~\cite{Thomson:2015jda, Lai:2021rko}. At higher energy, di-Higgs, top quark pair and $t\bar{t}H$ production lead to six-jet and even eight-jet final states and jet clustering becomes the dominant experimental limitation~\cite{Boronat:2016tgd}. Improved algorithms can have a profound impact on the potential to measure e.g. the Higgs self-coupling.

Machine-induced backgrounds at $e^+e^-$ colliders are generally benign compared to the pile-up levels encountered at the LHC, but can have a non-negligible impact on jet reconstruction, especially at higher energy. The VLC algorithm~\cite{Boronat:2014hva} modifies the beam distance criterion of the generalized $e^+e^-$ $k_t$ algorithm and has been shown to provide a much more robust performance in comparison to the classical sequential clustering algorithms for $e^+e^-$ collisions~\cite{Boronat:2016tgd} in the presence of $\gamma \gamma \rightarrow $ background. A thrust-based algorithm is found to yield better performance than $e^+e^-$ $k_t$ in two-jet events at the CEPC~\cite{Lai:2021rko}. The XCone algorithm~\cite{Stewart:2015waa} can naturally accommodate the boosted and resolved regimes and provides a close connection to calculations in Soft Collinear Effective Theory.

\paragraph{Muon colliders}

Since the major advantage of a muon collider is the ability to reach higher $\sqrt{s}$ than electron-positron colliders in a smaller area, the muon collider will produce final states that are generally more complicated than electron-positron colliders. Like other lepton colliders, the $z$ position of the interaction is also known precisely, and there is no pileup as in hadron colliders. As such, it is expected that jet algorithms developed for electron-positron colliders should also apply well to muon colliders. However, due to typically higher energies, boosted topologies tend to be more prevalent. 

\paragraph{Hadron colliders}

The challenges of a jet reconstruction at a hadron collider are well-known and extremely well-studied.
While jet substructure reconstruction and tagging techniques were not directly considered in the design of the initial detectors at LHC and their reconstruction algorithms, they provide excellent performance after several years of evolution in algorithms. For future hadron colliders, jet substructure reconstruction is already considered in their design and it is expected that similar techniques as currently deployed at the LHC will find broad applicability. 

The experiments at the LHC rely mostly on jets with a fixed distance parameter, where mostly the 
anti-$k_t$ algorithm with $R=0.4,0.8,1.0$~\cite{Cacciari:2005hq,Cacciari:2011ma} are used. The rigidity of the jet boundaries 
helps in pileup mitigation with an area-based 
approach~\cite{Cacciari:2008gn, CMS-PAS-JME-14-001, Aad:2015ina} 
and the calibration of isolated jets~\cite{Aaboud:2017jcu, Khachatryan:2016kdb}.
The experiences from the LHC have allowed extremely precise determination of jet energy and mass scales and resolutions (a review can be found in Ref.~\cite{Kogler:2018hem}), and have overall excellent precision. 

While a larger value of $R$ reduces hadronization corrections in jet $p_T$ which scale 
as $1/R$, the influence of pileup and the underlying event increases with $R^2$~\cite{Dasgupta:2007wa}. 
At the LHC experiments, it was possible to generally balance these effects with a few fixed-$R$ algorithms, variable-$R$ (VR) algorithms~\cite{Krohn:2009zg,Lapsien:2016zor} provide a promising alternative at future hadron colliders, which will have a larger dynamic range of jet energies, and have already been investigated by the LHC experiments~\cite{ATL-PHYS-PUB-2016-013,ATL-PHYS-PUB-2017-010,CMS:2020poo}. 


Particle-flow algorithms~\cite{ATLAS:2017ghe,CMS:2017yfk}, or more generally algorithms combining tracking and calorimeter information~\cite{ATLAS:2017ycq} are the state-of-the art to reconstruct jet substructure with fine granularity and good energy resolution.
Reconstruction challenges faced at the LHC, such as events containing up to 50 pileup interactions~\cite{CMS:2020ebo, ATLAS:2017ghe}, jet substructure of highly boosted $W$/$Z$/top quarks with multi-TeV transverse momenta, have been overcome successfully~\cite{CMS:2014joa, CMS:2020poo}.

\subsubsection{Jet substructure}

\paragraph{Electron-positron colliders}

Boosted object reconstruction at electron-positron colliders has been studied in full simulation by the CLIC group~\cite{Strom:2020wrm, CLICdp:2018esa}, with a focus on boosted top quark tagging. This study confirms the excellent response of the CLIC detector concept for a large number of substructure observables. 

In the electron-positron collider program at $\sqrt{s} \sim$ 250~GeV, jet substructure finds applications in many measurements. A good example is the measurement of the Higgs coupling to gluons, where the differences between quark and gluon jets can be used together with flavour tagging information to distinguish the $H \rightarrow gg$ decay from $H\rightarrow b\bar{b}$ and $H\rightarrow c\bar{c}$. Jet substructure observables and grooming techniques are likely of value in determinations of the strong coupling $\alpha_S$. This area has been identified as one of the open questions~\cite{Fujii:2020pxe}, but so far detailed phenomenology and experimental studies are lacking.
A lepton collider also offers excellent opportunities for jet substructure measurements that can be used to develop a better understanding of fragmentation and hadronization. Carefully validated first-principle calculations and models for Monte Carlo generators in the clean $e^+e^-$ without QCD radiation in the initial state can be very valuable in the preparation of a high-energy hadron collider.

\paragraph{Muon and Hadron colliders in the multi-TeV regime}

Boosted object reconstruction and tagging is crucial at a muon collider~\cite{Delahaye:2019omf} or advanced linear collider~\cite{ALEGRO:2019alc,Bai:2021rdg} operated at a center-of-mass energy in the multi-TeV regime, in addition the clear needs at future high energy hadron colliders. 

The JSS tools developed in the last two decades provide a very solid baseline for the developments for a future 100 TeV hadron collider. However, in such high energies there are additional challenges to be met in both the detector design (discussed in Section~\ref{sec:detector_100tev_pp}) and in the development of the algorithms. First, the physics program at 100 TeV requires both standard model measurements with high precision using boosted objects in a \pt-regime similar to the one at the (HL-)LHC ($~\sim$0.5--2 TeV), while in parallel explore the energy frontier with ultra relativistic particles with momenta up to $\sim$10-15 TeV. Particularly in this \pt-regime, the decay products from the heavy objects result in hadronic jets that overlap significantly and are extremely difficult to reconstruct and explore the internal jet structure. It is therefore critical to have sufficient detector granularity in future colliders to sufficiently reconstruct JSS. 

In addition to detector considerations, one algorithmic approach followed to overcome this challenge is to use only track-based variables for the design of the JSS algorithms, exploiting the much finer granularity of the tracking system compared to the calorimeters. These will be discussed in detail in Section~\ref{sec:trkobs}. 

\subsubsection{Mitigating beam backgrounds}

As noted above, beam backgrounds from electron-positron colliders is relatively benign compared to muon and hadron colliders though techniques have been developed to account for them.  

\paragraph{Muon colliders}

At a high-energy muon collider, interactions of the decay products of the muon beams with accelerator and detector elements create an intense flux of particles through the experiments. While the background screening and mitigation strategy is still under development, it seems likely that the residual background level~\cite{Collamati:2021sbv,Bartosik:2021bjh} requires a combination of active background mitigation in the low-level reconstruction algorithm and the use of robust high-level reconstruction algorithms. Examples of possibly effective low-level reconstruction techniques could include vertex association, tracklet pointing, and timing information. 

\paragraph{Hadron colliders}

Pileup mitigation is an important aspect of the low-level calibration of 
calorimeter cells, as pileup adds a diffuse noise term with large fluctuations. 
High granularity of the calorimeter is a pre-requisite for the mitigation of these 
effects, as it ensures an unambiguous combination with information from the tracking 
detectors. Machine learning techniques can help to improve the jet resolution by
identifying electromagnetic deposits within jets, which are then calibrated to the 
electromagnetic scale. This can lead to an improvement in the single particle 
response by about 50\%~\cite{Aleksa:2019pvl}, but the existing studies need to be extended to more realistic conditions including pileup and electronics noise. 

Jet reconstruction at the LHC is complicated by pileup. Pileup impacts jet reconstruction in a variety of ways, creating additional jets, changing the jet energy scale, and smearing out the jet energy resolution. It is particularly detrimental to jet substructure reconstruction, which can be affected by the presence of low-$p_T$ pileup particles. LHC experiments use a combination of several strategies to reduce the impact of pileup, which have enabled high-quality jet substructure taggers and measurements, even under high pileup conditions. More study is needed to understand the impact of pileup on future hadron colliders, such as the FCC-hh and the SPPC, but the prospects for these colliders can be informed by the performance at the LHC.


Experiments at the LHC rely on a variety of different techniques to reduce the effect of pileup on jet reconstruction, including the topocluster reconstruction~\cite{ATLAS:2016krp}, particle flow using the primary vertex association for tracks~\cite{ATLAS:2017ghe,CMS:2017yfk}, Constituent Subtraction~\cite{Berta:2014eza}, SoftKiller~\cite{ATLAS:2020gwe}, and the Pileup Per Particle Identification (PUPPI) algorithm~\cite{Bertolini:2014bba,Cacciari:2014gra,CMS:2020ebo}. For jet substructure reconstruction, grooming algorithms also provide some amount of pileup suppression, in addition to the other benefits they provide. 

At the HL-LHC, pileup conditions will become even more challenging, with an average number of interactions per bunch crossing of around 200. Nevertheless, both ATLAS and CMS expect to maintain good performance, making use of detector upgrades and advanced reconstruction algorithms, based on studies of small-$R$ and large-$R$ jet reconstruction~\cite{ATLAS-CONF-2017-065,CMS_DP2021_001}. Detector upgrades will also enable studies on the use of timing detectors~\cite{ATLAS-TDR-31,Butler:2019rpu} and, in the case of CMS, a high granularity calorimeter~\cite{CMS:2017jpq}. Existing pileup mitigation algorithms will become even more important for jet reconstruction, and novel methods for pileup mitigation are also being explored, such as machine learning to improve pileup identification and subtraction~\cite{Komiske:2017ubm,ArjonaMartinez:2018eah,Mikuni:2020wpr,Maier:2021ymx}.

Pileup conditions at the FCC-hh are expected to reach around five times those of the HL-LHC, with up to 1000 simultaneous proton-proton collisions. With this density of interactions, high quality spatial and timing resolution will be critical in order to resolve the different pileup vertices and associate tracks to them. ATLAS and CMS both rely on vertex association of tracks to reduce pileup for particle flow algorithms, and in the case of CMS, for the PUPPI algorithm. This means that the use of 4D tracking will be critical for jet substructure reconstruction at the FCC-hh/SPPC. While charged particles are able to provide useful inputs to jet substructure reconstruction, neutral particles provide additional information that can be used to improve the performance of jet taggers. To use this information effectively will require advances in particle flow reconstruction in dense environments as well as dedicated pileup mitigation algorithms. The HL-LHC will enable critical studies of new tools which can be used to reduce pileup effects at future colliders like the FCC-hh and SPPC, such as the use of timing detectors for object reconstruction, as well as the development of pileup mitigation algorithms for reconstructed inputs.

\section{Enhancing Sensitivity}
\label{sec:enhance}

In this section, we highlight applications and techniques for using JSS information to enhance the sensitivity of both measurements and searches at colliders.  First, we discuss novel and more exotic signatures of JSS which illustrates the broader application of the techniques we have discussed to search for potential new physics.  Then, we will describe a number of important and emerging techniques for analyzing JSS information.  In both cases, we cannot cover all approaches as JSS techniques are continually evolving in novel applications.  Instead, we present here a broad set of examples to give the reader a sense of the possibilities.  In addition to the direct physics possibilities, JSS serves as a test bed for new and creative ideas in theory and analysis.  The following section titles do not uniquely categorize the examples, which could be classified in a variety of ways.

\subsection{Uncovered Scenarios}


Traditional event reconstruction is mostly based on the principle that physics objects of interest can be individually reconstructed and well-isolated from other objects.
However, SM and BSM signatures can give rise to highly collimated objects, manifesting in unusual topologies which are relatively rare at the (HL-)LHC, but will be much more prevelent at future colliders.
Unconventional signatures can include cases where jets are composed of leptons and hadrons, only leptons, only photons, hadrons and missing transverse energy etc.  In addition to the jet kinematics and JSS, the jet timing~\cite{Chiu:2021sgs} information and other information can be used for classification. Examples include jets containing one or more hard leptons~\cite{Chatterjee:2019brg,Mitra:2016kov,Nemevsek:2018bbt,duPlessis:2021xuc,Dube:2017jgo}, displaced vertices~\cite{Nemevsek:2018bbt}, hard photons~\cite{Wang:2021uyb,Sheff:2020jyw}, or significant missing transverse momentum~\cite{Kar:2020bws,Canelli:2021aps}. Some of these anomalous signatures are already started being explored at the LHC~\cite{CMS:2021dzb,ATLAS:2019isd,CMS:2021dzg,ATLAS:2019tkk,CMS:2019qjk}.
Timing information can be useful to gain sensitivity in the searches with delayed jets~\cite{Liu:2018wte}. It will also enhance the accuracy of prompt jet and MET reconstruction, that can boost the sensitivity of several new physics searches. Moreover other detector upgrades for high radiation tolerance, unprecedented granularity particularly in the forward region~\cite{CERN-LHCC-2017-023,Allaire:2018bof}, extension of the detector acceptance~\cite{CERN-LHCC-2017-021,CERN-LHCC-2017-005,CERN-LHCC-2017-009}, a significantly sophisticated design upgrade of the trigger system~\cite{CERN-LHCC-2020-004} etc., will effectively lead us to broaden the search corners.

\subsubsection{Photon Jets} 

Axion-like particles (ALPs) are predicted by several extensions of the SM (e.g. spontaneous breaking of a global symmetry, hierarchy problem, an interesting connection to the puzzle of dark matter). The discovery potential of ALPs in the future LHC era can well be estimated in the mass range of ALPs, which is inaccessible to previous experiments~\cite{Wang:2021uyb}. The jet kinematics and a few JSS variables (e.g. hadronic energy fraction of a jet, number of charged tracks in a jet, N-subjettiness, fraction of the jet p$_T$ carried by the leading subjet, energy correlation function of the three hardest subjets) or the jet image study based on CNN technique~\cite{deOliveira:2015xxd,Ren:2021prq} are found to be extremely useful to disentangle photon-jet events from the single photon or QCD events. The so-called photon jet can be produced from the decay of boosted ALPs in the HL-LHC period. A detailed study of the reconstruction of a photon-jet, its calibration and performance in the future LHC environment or beyond needs to be carefully undertaken.

\subsubsection{Delayed Jets}

Several BSM predictions (e.g., supersymmetry (SUSY) with gauge-mediated SUSY breaking, hidden valley models, a Higgs boson decaying to glueballs where the Higgs boson is the portal to a dark QCD sector whose lightest states are the long-lived glueballs) lead to the unusual signature of non prompt or delayed jets which are sensitive to the proper measurement of jet timing~\cite{Chiu:2021sgs}. These non prompt or delayed jets are usually modeled to be produced by the displaced decays of the heavy long-lived particles in BSM. The sensitivity of these long-lived particle searches using non prompt or delayed jets is found to be significantly enhanced by the precision timing information of the jet. The time profile of a jet can be used as an independent probe of jet properties.  Similar to a choice of a jet clustering algorithm, the choice of a jet timing definition determines its properties and performance. The evaluation of various jet timing definitions is carried out depending on the closest representation to the parton level information as well as on the basis of minimizing the spread in the arrival times of the particles. Among the various jet timing definitions studied, the definition based on the p$_T$ weighted sum of the arrival times of the jet constituents exhibits the most promising performance both for prompt and delayed jets. However, the jet timing performance of a prompt jet is estimated to depend on its $\eta$ whereas the jet timing performance of a delayed jet is sensitive to the full kinematics of the event.   

\subsubsection{Dark QCD}

Searches for dark matter (DM) particles in colliders have remained unsuccessful so far. 
Consequently in recent years, some focus has shifted to unusual final states, which are not covered by typical searches at the LHC.
Semi-visible jets~\cite{Cohen:2015toa,Cohen:2017pzm} arise in strongly interacting dark sectors, where parton evolution includes dark sector emissions, resulting in jets overlapping with missing transverse momentum. This signature is usually discarded in the experiments, as it is usually from mismeasured jets. The implementation of semi-visible jets is done using the Pythia Hidden valley module~\cite{Carloni:2010tw,Carloni:2011kk} to duplicate the QCD sector parton shower. In a study~\cite{Kar:2020bws}, several jet substructure observables have been examined to compare semi-visible jets (signal) and light quark/gluon jets (background). The focus was on the more challenging scenario of $t$-channel production mode of semi-visible jets, where the absence of a resonance mass peak makes identifying the substructure difference more critical. The key parameter in the mode is the 
ratio of the rate of stable dark hadrons over the total rate of hadron, denoted by $R_{\mathrm{inv}}$. In general, it was found that $D_2$~\cite{Larkoski:2014gra}, $C_{2}$~\cite{Larkoski:2013eya} and ECF2~\cite{Larkoski:2013eya} observables were highly sensitive.
The overall interpretation is that the semi-visible jets result in more multi-pronged substructure. This was verified by clustering stable dark hadrons in the jets, which resulted in the differences in the substructure observables disappearing. This indicate that the substructure becomes less two-pronged with visible and dark hadrons in them, and the absence of the dark hadrons create the two-pronged structure. Detailed studies of this phenomenon can be found in Ref.~\cite{Cohen:2020afv}. 




\subsection{New observables}
\label{sec:newobs}

As more information is obtained in the realm of JSS, new observables can be constructed that have interesting properties, either experimentally or theoretically. We discuss a few examples in this section. 

\subsubsection{Track-based observables}
\label{sec:trkobs}

One of the challenges of an extremely high energy collider, such as a 100 TeV proton-proton collider, is so-called ``hyper-boosted'' jets, whose decay products will be collimated into areas the size of single calorimeter cells~\cite{larkoski2015tracking}. This fact, coupled with additional contamination from excess radiation - pileup, ISR, FSR, and UE, means that current jet substructure approaches will not be sufficient at a future high-energy hadron collider. One of the proposed mitigation strategies is to use track-based observables to augment calorimetric information~\cite{larkoski2015tracking,trackbased2013,track-assistedmass2019, trackingspannowsky2015}.

Studies have shown~\cite{larkoski2015tracking,ATLAS-CONF-2016-035,Gouskos:2642475} that this improves the identification performance in the ultra-relativistic limit. On the other hand, this result in imperfect measurement of the mass of the jet. Simple mass re-scaling techniques, e.g., as in Ref.~\cite{larkoski2015tracking,ATLAS-CONF-2016-035}, or more sophisticated ML-based and other approaches utilized at the LHC~\cite{ATL-PHYS-PUB-2017-015,CMS-DP-2021-017}, provide promising solutions to improve the mass reconstruction. However, a calorimetry system with sufficient granularity can be very important in JSS, even at the ultra-relativistic regime, as detailed in~\cite{calohep}. Based on these results, calorimeters at 100 TeV should have $\mathcal{O}$(10) finner granularity than the LHC calorimeters.

One study aimed to apply these strategies to the identification of hyper-boosted top quark jets~\cite{larkoski2015tracking}. First, the jet radius was scaled inversely with $p_T$ in order to remove excess radiation. While calorimetric information was then still sufficient to measure jet energy, tracking information was added in order to resolve substructure information, including the following track-based observables:
\begin{itemize}
    \item \textbf{Jet mass: $m = \frac{p_T}{p_{T}^{tracks}}m_{tracks}$}. The track-based mass $m_{tracks}$ is scaled in order to recover the neutral particle information that is not measured by the tracker.
    \item \textbf{Prongy-ness variables: $N$-subjettiness~\cite{N-subjettiness2011} and $n$-point energy correlation functions~\cite{ECF2013}}. These variables measure the likelihood of a jet to have a given number of subjets, and in this case only include track information. 
\end{itemize}
Clear improvement in the identification of top quarks vs both light quark jets and gluon jets is seen when using these track-based observables, as compared to calorimeter-based observables.

In~\cite{trackingspannowsky2015}, it is shown that the HEPTopTagger~\cite{heptoptagger1_2010,heptoptagger2_2010} can be successfully modified with a track-based approach (called HPTTopTagger) in order to identify tops at a future hadron collider. This technique is also applied to extremely boosted hadronic $W$- and $Z$-tagging, with the so-called HPTWTagger and HPTZTagger, respectively. Additionally in~\cite{topTagEnergyFrontier2018}, it is pointed out that these track-based observables could be further enhanced by so-called ``tracking calorimeters'', in which detailed information about individual particle decays could be reconstructed~\cite{detector100TeV2017}.

\subsubsection{$Flow_{n,5}$}
Taking into account the extremely collimated nature of heavy object jets, the quantities $Flow_{n,5}$ are introduced in~\cite{mangano2016physics}:
\begin{equation}
    Flow_{n,5} = \sum_p\frac{|p_T^p|}{|p_T^{jet}|}\,,
\end{equation}
where $n$ goes from 1 to 5 and $p$, $p_T^p$, and $p_T^{jet}$ are the jet constituents, jet constituent transverse momentum and jet transverse momentum, respectively. The sum runs over the jet constituents so that the following holds:
\begin{equation}
    \frac{n-1}{5}R \leq \Delta R (p,jet) < \frac{n}{5}R\,,
\end{equation}
where $\Delta R (p,jet) = \sqrt{(\Delta \eta)^2 + (\Delta \phi)^2}$ is the angular separation between the jet axis and a particular jet constituent, and $R$ is the jet size. These variables are applied to distinguishing boosted hadronically-decaying Z bosons from Randall Sundrum graviton decays ($G_{RS} \rightarrow ZZ$) to light quarks originating from $G_{RS} \rightarrow q\bar{q}$, specifically in the case where the $G_{RS}$ mass is equal to 32 TeV. It is found that the combination of jet mass and the $Flow_{n,5}$ variables outperforms the combination of jet mass and the $\tau_2/\tau_1$ $N$-subjettiness ratio.

\subsection{Novel Physics Effects}

New showering effects begin to emerge at multi-TeV energies, including gluon splitting to top quark pairs, weak bosons radiating from jets, and radiation off of top quarks. These can affect boosted object identification overall, but these particularly affect boosted top quark identification, because they correspond to real on-shell weak bosons or top quarks, or enhanced radiation off of quarks. These effects are explored in the case of boosted top quark identification~\cite{topTagEnergyFrontier2018}.


At very high energies, a gluon can split directly into a top quark pair~\cite{topTagEnergyFrontier2018}. This phenomenon will therefore increase gluon mistag rates. To mitigate this affect, it is important to recognize the fact that a gluon jet will have more constituents than a prompt top quark, and so the gluon-induced top will carry a smaller percentage of the jet's total energy. A useful discriminating variable would therefore be the transverse momentum ratio of a top-tagged subjet to its host large-radius jet: $p_{T, \text{top-subjet}}/p_{T, \text{fatjet}}$.

At extremely high energies, particles will radiate $W$, $Z$, and $h$ bosons~\cite{topTagEnergyFrontier2018}. This can lead to light quarks jets that look like heavy particle jets. In the case of semi-leptonic top-tagging, $W$-strahlung can be particularly problematic. To mitigate this problem, one can take advantage of the fact that $W$-strahlung emissions peak at an angle of about $5m_W/p_T$, whereas a top decay happens within a cone of approximately $m_t/p_T$~\cite{semileptonicTopWstrahlung2011}. Therefore, upper bound on the angle between the $b$-jet and the muon can be used to discriminate between tops and light quarks that radiate $W$ bosons. Further study is required to mitigate the effect of weakstrahlung on other heavy particle tagging scenarios.
Additional kinematic handles and AI/ML-based techniques may be deployed to provide further discrimination.

Similarly, identifying $WV\rightarrow\ell\nu qq$ from heavy particle decays is an important but challenging problem due to overlapping lepton and jet signatures~\cite{Agashe:2018leo}. ML-based taggers, such as convolutional (CNN) and/or fully connected (DNN) trained to distinguish signal (boosted $WV$) and background (QCD multijets) based on calorimeter and tracking features in jet constituents, can be used to enhance sensitivity to new physics in future hadron colliders. 

Even after applying a shrinking jet radius, a reconstructed top jet will still include some semi-hard final state radiation. This leads to around 10\% of these jets having obfuscated substructure and masses well above the top mass. In this case, one can improve top/gluon discrimination by treating top quarks similarly to light quarks, and take advantage of the fact that a top will have less wide-angle radiation. As a simple example, it was shown that adding a track counting variable to a top tagger could improve discrimination, reducing gluon mistag rates by up to 20\%~\cite{topTagEnergyFrontier2018}.

\subsection{Novel analysis techniques}
\label{sec:noveltechs}

The theoretical and experimental innovation discussed above will also require novel analysis techniques. These cannot be entirely of a computational nature, but will also require re-imagining the inputs to JSS and their processing. Selected examples on iterative generator tuning, anomaly detection, and machine-learning assisted techniques are outlined below.

\subsubsection{Iterative Monte Carlo generator tuning}
JSS techniques are sensitive to simulation effects such as underlying event and parton shower modeling, see also Sec.~\ref{sec:mctuning}. While this directly affects the sensitivity of physics analyses that are making use of JSS techniques, it provides also the opportunity to constrain and improve physics modeling by performing dedicated measurements. In the past, these measurements have been performed in a one-off manner, i.e.\ the measurements are published~\cite{CMS:2018ypj,Aaboud:2019aii}, then used for a future tuning campaign~\cite{CMS:2020dqt,CMS:2021byj} in a systematic way~\cite{Buckley:2009bj}. This approach, however, integrates over a huge phase space and often yields suboptimal values for JSS~\cite{CMS:2020dqt}. With the help of declarative and therefore consistently repeatable workflows~\cite{declarativewfs} and machine-learning techniques, this approach can be significantly improved. The generator settings can be adjusted iteratively by repeating dedicated JSS measurements, consequently yielding optimal settings for the given suite of measurements. This can similarly be achieved by using machine-learning techniques to determine optimal generator settings (see e.g., Ref.~\cite{1907.08209}), for example to minimize related uncertainties. By adding further measurements, also not those directly related to JSS, significantly better simulation can be achieved.  Enhanced tuning (and also a variety of measurements) may be enabled by unbinned and high-dimensional differential cross section measurements that are not possible by ML (see e.g. Ref.~\cite{Arratia:2021otl}).

\subsubsection{Anomaly detection}
One of the most promising applications of machine-learning in ATLAS and CMS could be model-agnostic anomaly searches. There are a vast number of interesting new physics scenarios that we would like to search for at the LHC, however using traditional hypothesis testing techniques it is not possible to search for all them. Anomaly detection techniques aim to circumvent this problem by automatically identifying potential BSM contributions.  These anomalies could be outliers (low probability density) or over/under-densities in phase space with respect to the SM.
In this approach, a specific signal hypothesis is not required, although there is a tradeoff between performance on a given scenario and model dependence.  Anomaly detection can be applied to individual objects/jets or to entire events.
Modern deep-learning techniques dramatically increase the sensitivity of anomaly detection methods through their ability to use low-level, high-dimensional inputs. 
The technical concept behind these new anomaly searches is unsupervised, weakly supervised, and/or semi-supervised training of deep classification networks (see Ref.~\cite{2010.14554,Kasieczka:2021xcg,Aarrestad:2021oeb,2112.03769} for recent reviews).

A notable application is in the use of autoencoder neural networks optimised to compress and reconstruct event data. The accuracy of the reconstruction can then be used as the observable with which to identify the anomalies for instance in jets~\cite{Heimel:2018mkt,Macaluso:2018tck}. Anomalous events may be expected to occur much less often in the data and thus result in less accurate reconstruction by the autoencoder. A promising path to improve this method is to extend the discriminative power from the physics phase space to include the latent space of the neural networks. This can be achieved, for example, using rapidity-mass matrices for standard autoencoders \cite{Chekanov:2021pus}, (Dirichlet) variational autoencoders~\cite{Dillon:2021aeo,Dillon:2021nxw} or invertible normalizing flow network~\cite{Buss:2022lxw}, benchmarked for dark-matter-inspired jet signatures. For any kind of neural network application to jet physics, self-supervised learning of symmetries,  fundamental invariances, and detector effects is an exciting new direction which is expected to significantly improve the understanding and the experimental stability of neural networks applied to subjet physics~\cite{Dillon:2021gag}.

Related applications of anomaly detection, such as the classification without labels (CWoLa) method~\cite{1708.02949}, are promising tools to enhance bump hunt analyses~\cite{1805.02664,Collins:2019jip}; this approach is also the first ML-based anomaly detection method to be applied to collider data~\cite{Aad:2020cws}. In Ref.~\cite{Kasieczka:2021xcg} the results of the LHC Olympics showcase many different methods on a resonant anomaly detection challenge. Recent developments have brought in a better understanding of these deep-learning techniques and new ideas for background estimation~\cite{Mikuni:2021nwn} and linearized explanations of decision classifiers~\cite{Agarwal:2020fpt,Mokhtar:2021bkf}. Ongoing and future work will certainly lead to more progress in all of these areas.

\subsubsection{Hit-based Inputs for High-Energy Flavor Tagging}
Studies are on-going at the ATLAS and CMS experiments to incorporate some of the ideas first explored in \cite{HitTaggerA}. When central jet energies exceed 500~GeV several effects make tracking difficult and the ability to discriminate jets containing B hadrons decreases. However, because a primary B hadron will often absorb most of the jet's energy, it has a high probability of crossing the innermost layer or layers of trackers in colliding beam machines prior to decay. Using the fact that hit patterns might ``jump'' from one layer to the next, or that charged tracks would cluster more tightly around the jet axis could be used as contributing input to sophisticated ML algorithms to improve their performance in the high energy regime. Initial studies are indicating that some additional efficiency can be gained up to 1500 GeV with hit-based inputs added to neural network-based taggers.\cite{HitTaggerB}. If found effective, this technique might influence tracker design at future colliders where high energy jets will be even more common than currently at the LHC.


 
 
\section{Executive Summary}

In lieu of conclusions, we offer a summary of jet substructure at future colliders. Jet substructure (JSS) has emerged as a powerful framework for studying the Standard  Model (SM) and provides a key set of tools for probing nature at the highest energy scales accessible by terrestrial experiments.  While not an experimental or theoretical consideration of the design of the original LHC experiments, JSS is now being widely used to extend the sensitivity of searches for new particles, to enhance the precision of measurements of highly-Lorentz-boosted SM particles, as well as to probe the fundamental and emergent properties of the strong force in new ways.  Along the way, the JSS community has been a catalyst for new detector concepts, new analysis tools (e.g., deep learning), new theory techniques, and more.  Jet substructure has been transformative for the physics program of the LHC and it can play a central role in the physics case for future colliders. 

\section*{\label{sec::acknowledgments}Acknowledgments}

BN was supported by the Department of Energy, Office of Science under contract number DE-AC02-05CH11231. GC acknowledges support by the Funda\c{c}{\~ a}o para a Ci{\^ e}ncia e a Tecnologia (Portugal) under project CERN/FIS-PAR/0024/2019 and contract 'Investigador auxiliar FCT - Individual Call/03216/2017' and from the European Union's Horizon 2020 research and innovation programme under grant agreement No. 824093. SR and CM were supported by the National Science Foundation under grant 2111229. AH gratefully acknowledges funding by the Deutsche Forschungsgemeinschaft (DFG, German Research Foundation) in the Emmy-Noether program (HI 1952/1-1 and HI 1952/1-2) and under Germany’s Excellence Strategy – EXC 2121 "Quantum Universe" – 390833306. DK is funded by National Research  Foundation (NRF), South Africa through Competitive Programme for Rated Researchers (CPRR), Grant No: 118515. BMD was supported by a Postdoctoral Research Fellowship from Alexander von Humboldt Foundation. BTH and RL acknowledge the support of STFC, United Kingdom. NT is supported by Fermi Research Alliance, LLC under Contract No. DE-AC02-07CH11359 with the DOE, Office of Science, Office of High Energy Physics and the DOE Early Career Research program under Award No. DE-0000247070.

\bibliographystyle{JHEP}
\bibliography{main}

\end{document}